\documentstyle{aipproc}

\def\etal{{\sl et al.}}

\def\d0{D\O} 

\def\h1{H1} 

\def\tbar{$\overline t$}
\def\ttbar{$t$\tbar}
\def\bbar{$\overline b$}
\def\bbbar{$b$\bbar} 
\def\cbar{$\overline c$}
\def\ccbar{$c$\cbar}

\def\eebar{$e^+ e^-$} 

\def\ifb{fb$^{-1}$}
\def\lumunits{cm$^{-2}$s$^{-1}$}
\def\lum{${\cal L}$}

\newcommand{\lsim}{\mathrel{\lower4pt\hbox{$\sim$}}
\hskip-14.5pt\raise1.6pt\hbox{$<$}\;}

\newcommand{\gsim}{\mathrel{\lower4pt\hbox{$\sim$}}
\hskip-13.5pt\raise1.6pt\hbox{$>$}\;}

\def\be{\begin{equation}}
\def\ee{\end{equation}}
\def\bea{\begin{eqnarray}}
\def\eea{\end{eqnarray}}

\begin{document}
\vspace*{4cm}
\title{Linear Collider Workshop 2000 Summary} 

\author{ P.D. Grannis }

\address{Department of Physics and Astronomy,\\
State University of New York, Stony Brook NY 11794}

\maketitle
\begin{abstract}
We summarize some of the main physics questions that will
serve to define the linear $e^+e^-$ collider program,
and comment on issues that confront the world community
in making such a collider a reality.
\end{abstract}

The 1999 International Workshop on Linear Colliders (LCWS) at Sitges, Barcelona
~\cite{sitges,richardsitges} demonstrated the wide range of physics
opportunities at a 500 -- 1000 GeV linear
\eebar collider (LC).  
Since that time, there has been considerable progress
on R\&D and technical proposals for a collider, 
and in further exploring
the physics potential.    The 2000 LCWS~\cite{lcwsfnal}
at Fermilab gave the opportunity to update the physics case and detector needs, 
and to review accelerator developments.

The linear collider proposals under discussion are
TESLA~\cite{napoly}, JLC~\cite{chin} and  NLC~\cite{raubenheimer}.
TESLA proposes superconducting cavities operating at 1.3 GHz and effective
gradients of 22 MV/m.  The JLC or NLC X-band proposals employ warm accelerating
structures at 11.4 GHz with 50 MV/m.  The JLC C-band machine would 
operate at 5.7 GHz and 34 MV/m.
Each envisions a first phase of operation at up to $\sqrt s =$ 500 GeV.
The luminosity (at 500 GeV)
expected for TESLA is \lum = 3.4 $\times 10^{34} $\lumunits; 
for the X-band JLC or NLC, \lum = 2.2 $\times 10^{34} $\lumunits; 
and for the C-band JLC, \lum = 4.3 $\times 10^{33} $\lumunits.
TESLA could be upgraded to about 800 GeV;  the X-band
JLC/NLC is upgradeable to about 1000 GeV with known technology.   
Recall that 
\lum = 1 $\times 10^{34} $ \lumunits gives 100 \ifb accumulated data 
for a `standard operating year' of $10^7$ s.
Future higher energy \eebar colliders may depend upon the use of high
intensity, low energy drive beams as a source of RF power; this concept
has been pioneered by the CLIC~\cite{ianwilson} R\&D program at CERN.

The superconducting TESLA machine and warm cavity JLC/NLC designs differ
in several important respects that could influence a final choice of collider
technology.  However, the physics case 
likely depends only weakly on this choice, once the maximum energy and
luminosity are fixed.   Thus, the work across the world to understand the
physics capabilities and delineate the parameters of detectors are relatively
interchangeable.   This commonality of the worldwide study of the 
linear collider program has been a strong unifying force for the community.

\vskip 6mm
\leftline{\large {\it 1. The Linear Collider physics program}}       
\vskip 1mm

The many contributions and plenary summaries at this workshop
make a strong and detailed case for the LC physics capability,
and I do not repeat this in depth here.   The range of physics
issues for LC study were summarized as part of the charge to
this workshop \cite{zerwas,komamiya}.   There has also been
a recent discussion of a first phase of the LC
at 500 GeV, outlining the case that substantial new physics will 
be accessible there \cite{uswhite}.

\vskip 3mm
\leftline{\it 1.1 Higgs studies}               
\vskip 1mm

The clearest and most pressing physics issue at the LC is 
the full study of the Higgs sector.   In both the Standard
Model (SM) and its supersymmetric extensions, we confidently
expect a Higgs boson
with mass below 200 GeV.  In most strong coupling models, or models
with large extra dimensions, Higgs-like objects are expected at a few
hundred GeV.   For all but some special corners of parameter space,
such states should be discovered, and the mass determined to
good precision at the LHC.  However the LHC will likely fail to fully delineate
the Higgs:~~branching ratios to fermions, $J^{PC}$ determinations, 
CP-violation properties, the total
width, and Higgs 
self-couplings are all difficult to determine, or are inaccessible, 
at the LHC.  In the case that the Higgs is embedded in supersymmetry,
further measurements of the other Higgs masses, and 
the ratio of up- and down-quark
couplings ($\tan\beta$) will also be difficult in a model-independent context.
Indeed, as we move away from the simplified supersymmetric paradigms to
consideration of the full supersymmetric model with over 100 soft parameters,
we will need many measurements in the Higgs (and sparticle) 
sector to learn what these
parameters are in Nature.    

The LC should be capable 
of making a full portrait of the Higgs bosons, given enough
energy to produce them \cite{desch,okada}.   The LC
will make only marginal improvements in the Higgs mass determination
over the LHC \cite{shapiro}.  However, it will allow measurement of the 
total width to a few percent, even for low mass Higgs where the LHC
has no capability due to finite energy resolution.   The total width is a key
parameter for sensing potential new decay channels, multiple Higgs states, or
non-SM effects.
Though the LHC can rule out spin 1 through observation of
$H \rightarrow \gamma\gamma$, and can provide some quantum number information
for high mass Higgs through the $ZZ$ final state angular correlations,
the LC can give full quantum number determination for any Higgs mass using
polarized beams and threshold scans.

Determination of the branching ratios of the Higgs to 
fermion pairs will be crucial for distinguishing the SM from 
supersymmetric or other SM extensions.   Studies
\cite{battaglia,desch} show that for $M_{Higgs}$ below 150 GeV, the branching 
ratios for SM Higgs decays into \bbbar, \ccbar, $gg$, $\tau^+ \tau^-$, 
$WW$ can be determined to 5 -- 10\%
with 500 \ifb .  The $\gamma\gamma$ branching ratio can be measured
for low Higgs mass ($< 120$ GeV) to perhaps 20\%.  
The Higgstrahlung and $WW$ fusion processes can be differentiated
to give information on the $g_{HWW}$ and $g_{HZZ}$ couplings.
Operating the LC above the \ttbar$H$ threshold will give the top Yukawa
coupling to the 10\% level.   Finally, the crucial self-couplings
of the Higgs bosons, key to understanding the character of the Higgs potential,
can be determined through the study of double Higgs bremsstrahlung;
these measurements will be statistics limited \cite{desch} and may require
very good jet energy resolutions.
The full set of possible Higgs measurements would
benefit from further study using complete realistic 
simulations of potential detectors.

The combination of all this high precision Higgs information can be used in
an analogous way to that of the LEP/SLC/Tevatron data constraining the SM.
Departures from SM orthodoxy for branching ratios, total width, or
couplings would be expected in any new physics model, so the Higgs sector
measurements will have high priority.  For example, the branching fraction
data at a 500 GeV LC is capable of sensing a supersymmetric pseudoscalar
Higgs ($A$) up to about 700 GeV \cite{battaglia}.  Understanding
the properties of a Higgs candidate discovered at the Tevatron or
LHC and seeking departures from the SM will form the backbone of
the LC program for several years, and gives a strong justification
for acquiring high luminosity samples.   This program will occupy the
initial years of LC operation if a Higgs candidate emerges below about
200 GeV.   The Higgs sector measurements also will require substantial
luminosity accumulations at energies at $\sim 1$ TeV 
to determine the $t\overline t H$
Yukawa coupling, and to explore the heavier Higgs states if supersymmetry
is part of the new physics.

\vskip 3mm
\leftline{\it 1.2 Studies of the supersymmetric partners}  
\vskip 1mm

If supersymmetry is responsible for electroweak symmetry breaking, we
confidently expect observable effects at the LHC\cite{shapiro}.  The LHC should 
be able to determine mass differences for particles in the cascade decays
of the strongly produced squarks and gluinos, and thus to measure
Susy model parameters given an assumption of the model class, sometimes
to percent level precision.  The LHC will however not typically be
sensitive to the heavy gaugino or Higgs states, will 
not see sleptons unless
they are prominant in squark or gluino decays, will typically not study the
sneutrinos, and will likely not be
capable of measuring sparticle quantum numbers, gaugino or sfermion mixing
angles, or probe the CP-violating phases.

The LC measurements of Susy particle properties and mixings, 
leading to determination
of the underlying Susy model parameters will extend the LHC studies
dramatically \cite{martyn,godbole}.   The ability in the LC to polarize
the electron (and perhaps positron) beam allows selection of particular
subprocesses of interest, or permits suppression of SM backgrounds.
The ability to set the parton cm energy above successive sparticle pair
thresholds will help disentangle backgrounds from Susy itself.  Precise
measurements at the LC will in several cases improve the utility of
the LHC results; for example, knowledge of the LSP mass from the LC will
allow LHC mass differences to be converted into masses.   The LC and LHC
are often complementary -- the LHC should access the sparticle
states that couple to color, up to very high mass, whereas the LC will excel
in studies of the slepton and gaugino sectors.

If the LHC shows that supersymmetry has a key role in EWSB, the next
question of paramount importance is how the supersymmetry breaking occurs,
and at what scale the symmetry is broken.  Though most studies of Susy are
now conducted in a specific simplifying framework such as gravity mediating
Susy, gauge mediation, anomaly mediation or
gaugino mediation with relatively few parameters,
the real model is likely to be considerably more complex.  
To some extent, the sparticle mass patterns can indicate the general
type of model chosen by Nature \cite{martyn}, but to fully 
realize the model more experimental information will be needed.  
We recall that the minimal
supersymmetric model contains over 100 arbitrary parameters, and that a full
understanding of Susy requires that we determine them all.  Typically, the
CP-violating phases are set to zero in simulation exercises, but it may well
be that non-zero phases are required to explain the CP violation patterns seen
in the $K$ and $B$ systems, and to explain 
the baryon-antibaryon asymmetry in the universe \cite{kane}.  Measurements
of CP-violation in supersymmetry may be the next new frontier in flavor physics!

Measuring a large set of Susy observables -- masses, mixing angles
in chargino, neutralino, stop and stau sectors,
branching fractions, Higgs sector parameters etc. -- will be necessary
to fully determine the Susy model.  It is likely that the measurement
of the Susy model parameters will form a productive 
interface to string theoretic models, giving experimental guidance
to Planck scale physics theory, and helping point the way to identifying
the intermediate scales of new physics between the TeV and Planck scales
\cite{kane}.  
Obtaining information on the multitude of new Susy parameters will require
extensive and precise measurements of as many new Susy particles as we
can produce.   Thus, the energy scale at which
we may be confident that the sparticles can be produced in the LC is
a critical question.   There are at present only plausibility arguments
to guide us.   If supersymmetry is to produce EWSB and yield the observed
$Z$ mass without `excessive' fine tuning, then it follows  that
we are likely to produce the lighter chargino and neutralinos, and the
sleptons, at a 500 GeV linear collider \cite{uswhite}.
If the LSP is to provide the candidate for dark matter in the universe,
light sparticles are preferred (though some of the parameter space
yielding Susy dark matter requires 
raising the LC energy to 1 TeV or a little higher \cite{ellis}).
If Susy is to provide the CP violation needed
for cosmic baryon asymmetries, light sparticles are favored.

However, in a model as rich and complex as supersymmetry, there are few
guarantees on where we will find the spectrum of new particles.  More to
the point, it is almost certain that some of them will be inaccessible
at a 500 GeV collider, and that to understand  the Susy symmetry breaking,
higher energy will be required.
Although finding supersymmetry will bring powerful justification for
a first stage of the LC at about 500 GeV, it seems likely that it will
also reinforce the case for energy upgrade to at least 1 TeV 
in a subsequent phase.

\vskip 3mm
\leftline{\it 1.3 Non-supersymmetric extensions to the Standard Model}
\vskip 1mm

Popular though supersymmetry is as a framework for going beyond the
Standard Model, 
there is at present no hard experimental evidence to support
this solution to electroweak symmetry breaking.  Indeed, in the past several
years alternative models for EWSB\cite{kim,kilian}
 have been elaborated in a way that
conforms to the precision EW data from current experiments, and which
possess strong intuitive appeal for many.  Thus it is important to
gauge the strength of the LC in exploring the consequences for such
models.   Two main non-Susy themes have been developed theoretically:
models in which there is a new strong coupling 
sector, typically at the
several TeV scale, and models in which the some of the string-inspired
extra spatial dimensions can be larger than the Planck length and enter
into the phenomenology of EWSB.

The strong coupling models emerged around 1980, based upon analogies
with QCD or superconductivity.  The effects of the SM Higgs boson 
could be mimicked by new physics at the TeV scale, often producing composite
Higgs-like particles.    The precision electroweak data accumulated over
the past 20 years place stringent constraints on the properties of such
theories, but viable models remain\cite{dobrescu}.   
The masses of the composite Higgs
states depend on the model envisioned, but are typically in the range
below 500 -- 600 GeV.   Its couplings to ordinary particles resemble
those of the SM, but new couplings arise that give observable effects.
In particular, anomalous $WW\gamma$ and $WWZ$ couplings are expected
in the range $\Delta \kappa$, $\lambda ~$ = few $\times 10^{-3}$ that should be
observable through $WW$ pair production in \eebar collisions, but would be
problematic for LHC.   Modification of the SM 
$Z\overline t t$ couplings would also be expected at the 5 -- 10\% level,
and these should be accessible at the LC using polarized electron beams.
Strong coupling theories typically modify $WW$ scattering for a LC
operating above 1 TeV, and at the LHC. 

Models incorporating large extra dimensions suggest many potential
new signatures, depending on which particles propagate into the bulk
and the size and number of extra dimensions\cite{arkani}.    In this space of
model types, many observable effects could come into play, including direct
observation of Kaluza Klein (KK) towers of gravitons that simulate contact
interactions, observable KK towers of gauge bosons resembling
excited $Z$ bosons, modifications to the 
angular and mass distributions of fermion or boson pairs,
or modification of scattering amplitudes at large momentum transfers.
Typically the LHC can probe new particles to somewhat higher mass, but the LC
can disentangle the character of the states more efficiently, and can
probe new physics in the lepton sector that is not available in hadron machines.

The unifying thread for the multitude of strong coupling and string
inspired models is the potential need for higher energy to directly access
the new physics.   
However, we have learned from the past that the constraints placed
by precision measurements at the 100 GeV scale have very substantially
limited the space of potential new models of these types.  Thus, if
future experiments suggest that it is in these directions that we must
go, the precision measurement arena will surely be important.
Samples of $10^9$ or more $Z$s should allow crucial improvements
of many of the electroweak parameters\cite{moenig}.    Operation of the LC
at the \ttbar ~and $WW$ thresholds will give great improvements in our knowledge
of the top quark and $W$ boson mass, and these are also
crucial for constraining new models.
Moreover, if there is a Higgs-like state to study, it too should be regarded
as fertile ground for precision measurements of its properties, as these
too give powerful guidance on the character of the new physics.
Thus, even in the case that we are in a realm of exploring 
non-SM physics whose new particle spectrum lies above
the initial energy of the collider, there are many critical measurements
at lower energies needed 
for delimiting this new world and for pointing to
the productive avenues for its exploration.

\vskip 3mm
\leftline{\it 1.4 Other physics opportunities }
\vskip 1mm

For most, elucidating the EWSB mechanism, and pointing the way to
physics at higher mass scales is the key element of the LC program.
There are however many other physics topics that will form important parts of
the LC program.   Some of these could be raised in importance if the particles
associated with EWSB are not clearly visible at the LHC or in the first stage of
the LC program.   As mentioned above, high statistics samples of $Z$'s may be
crucial to constrain new physics.   Improved precision of the $W$ boson
and top quark masses can be obtained from runs at the appropriate pair
thresholds, and these too may be a key ingredient in unravelling the
puzzle if Susy particles are not seen.    A large sample of $Z$ bosons
(${\cal O}(10^{10}$) would give impressive samples of clean \bbbar ~events
with which to pursue studies of the CKM matrix parameters, CP -violation,
and rare $b$ decays.

Electroweak measurements of $W/Z$ and \ttbar ~final states can probe
for new short distance effects.  The anomalous coupling studies of
$W$ and top will be more sensitive than those available from the LHC.
Large samples of \ttbar ~can be used to reconstruct the spin structure
of the interaction and probe for CP-violating effects \cite{orr}.

As has been true at lower energy \eebar colliders, the LC will enable
many precise tests of QCD and hadronic structure \cite{magill}.   
The precision
to which the strong coupling constant can be determined is estimated
at 1\%, using a variety of established methods.   The LC will open up
new opportunities for the study of $\gamma \gamma$ scattering and photon
structure.  The study of low-$x$ resummed BFKL effects should be particularly
clean, and will offer a new opportunity to understand QCD dynamics in
non-perturbative regimes.  Gluon radiation in top quark production and
decay offers new windows for probing non-perturbative effects in QCD.
Studies of \ttbar ~and single top quark
production in $\gamma\gamma$ collisions should complement
those in \eebar collisions.

Perhaps the most likely areas of physics beyond the usual EWSB topics
that will occupy us at the LC are those that are as yet dimly, if at
all, in view.  It has been refreshing in the past few years to see
the wealth of new speculative models suggesting observable phenomena 
at the LC.   It is clear that the phenomenology associated with large extra
spatial dimensions will grow richer in the near future.  At this meeting,
we have heard exotic suggestions \cite{kim} regarding new contributions
to the general relativistic metric (`torsion'); many variants
of the sets of particles propagating in extra dimensions;
suggestions of non-commutative field theories with observable consequences;
modifications to the equivalence principle; and 
search for `maximal weirdness' (quite likely inaptly named, since
theorists may be counted on to outdo themselves in weirdness every year or two).
The point of course is not that any of these conjectures has a high probability
to be realized in Nature, but that there will be a large variety of 
new ideas, and that the LC will have unique capabilities
to test them.

\vskip 3mm
\leftline{\it 1.5 Physics summary }
\vskip 1mm

The physics opportunities at the LC are rich and varied.   Although
we cannot state with absolute certainty that the source of EWSB will
be accessible at the 500 GeV LC, it is an extraordinarily good bet.
The SM Higgs can be studied in depth.  Supersymmetry is open for
study, both in the Higgs sector and at least the lighter gauginos
and sfermions.  Observable effects from new strong interactions, or
from models with large extra dimensions, are found in virtually all
existing models.   In any case, the observations at 500 GeV should
clearly point the way for upgrades in collider energy.

We should not expect that the LC will necessarily discover the new
phenomena related to EWSB.  
The Higgs or supersymmetry will likely 
be first seen at hadron colliders.
But we should not take discovery
of the new physics as the main criterion for deciding to build
a linear collider.   One may argue that LEP/SLC provided no
new `discoveries' (the demonstration of three neutrino species
with mass below $m_Z/2$ and normal $SU(2)\times U(1)$ quantum
numbers is an arguable exception).  Yet there can be no argument
that these experiments have dramatically and fundamentally altered
our understanding of particle physics.  Similarly, we are confident
that the LC will provide us with a true understanding of what the 
Higgs boson is, will delineate the nature of the supersymmetric world 
and point to its symmetry breaking characteristics, or will expose
the nature of strong coupling -- even if these phenomena are first
observed at the LHC.  For each of the broad cases studied so far,
the LC is essential to understand the new physics beyond the SM.

\vskip 6mm
\leftline{\large {\it 2. Experimental issues}}
\vskip 1mm

There are many issues related to the collider operating conditions
or the detectors that require fuller discussion in the worldwide
community.
Some of these are affected by the specific physics scenario that
we find ourselves in, so flexibility should be retained so
that later incorporation of options can be made.

\vskip 3mm
\leftline{\it 2.1 Positron polarization }
\vskip 1mm

The need for electron polarization has long been recognized; 
$|{\cal P}_{e^-}| = 0.9 $ could be achievable using improved
strained GaAs targets.  Positron polarization might be achieved using
pair production from polarized photons from undulator magnets, or from
backscattering off high power lasers; $|{\cal P}_{e^+}| = 0.5 $ seems
a sensible goal.

Positron polarization would give advantages similar to those from electron
polarization -- the ability to suppress backgrounds and enhance specific
signal subprocesses.  Examples from Susy include \cite{martyn}:
\begin{itemize} 
\item
The reactions
$e^+e^-\rightarrow \widetilde \chi_i^+ \widetilde \chi_i^- ,
~~\widetilde \ell \widetilde \ell$ can be dialled from mostly
$\widetilde \ell_R \widetilde \ell_{L/R}$ to dominantly 
$\widetilde \ell_L \widetilde \ell_L$ and 
$\widetilde \chi_i^+ \widetilde \chi_i^- $ as ${\cal P}_{e^+}$ is varied
from $-$0.6 to +0.6.   
\item
The precision with which $\widetilde t$ mixing
can be measured is enhanced noticeably using positron polarization.
\item
The polarization dependence of 
$\widetilde \chi_1^0 \widetilde \chi_2^0$ differs between minimal Susy
and extended models.   
\item
Allowing positron polarization gives improved
ability to measure mixings in the gaugino sector, and thus is a key
contributor to analyses that determine the nature of Susy breaking mechanisms.
\end{itemize}

The Higgs pair production cross section, used to measure the Higgs potential
parameters, is enhanced by up to a factor of two with positron polarization
\cite{desch,okada}.
If much better precision on $\sin\theta^2_W$ is required, new measurements
of $A_{LR}$ at the $Z$-pole will be needed; the errors arising from uncertainty
on beam polarizations are greatly reduced if the positron beam can also
be polarized\cite{moenig,kilian}.

The need for positron polarization as a tool to increase precision
and to disentangle rival processes is sufficiently widespread that
it is highly desirable that such a capability be allowed by the
design of the accelerator, even if it is not implemented at the very beginning.

\vskip 3mm
\leftline{\it 2.2 $\gamma \gamma$ collisions }
\vskip 1mm

$\gamma \gamma$ (or $e \gamma$) collisions can be
made by backscattering high power laser beams from the electron beams
\cite{takahashi,telnov};
polarized photon beams at about 80\% of the primary electron
beam energy can be produced.   The $\gamma\gamma$ luminosity is
about 0.4 times that for \eebar, and is peaked within about
15\% of the maximum energy.
Recent developments in lasers are
promising, but considerable work remains to develop the appropriate
mirrors, beam masks etc.   Special care must be given in detector
design to accomodate the high flux of \eebar halo backgrounds;  

The need for $\gamma\gamma$ collisions, like that for ${\cal P}_{e^+}$,
ranges widely over the potential LC program.   The total width of the Higgs
boson is a key measureable, and if the Higgs mass is below 200 GeV, is
inaccessible for the LHC.   Measurement of $\sigma(\gamma\gamma\rightarrow H)$
and $BR(H\rightarrow\gamma\gamma)$ should give the total width error
of 5\% (in 200 \ifb ).   Higgs production from circularly
polarized $\gamma\gamma$ collisions allows tests of CP violation in
the Higgs sector.    $\gamma\gamma \rightarrow$ Susy $H/A$ will help
distinguish the scalar and pseudoscalar states.    Chargino production
in $\gamma\gamma$ collisions  offers clean determination of gaugino
mass matrix parameters.   In many cases, new particle production
is enhanced in $\gamma\gamma$ collisions relative to \eebar.

Once again, we should foresee that some of the physics that evolves
at the LC could be enhanced strongly by the use of $\gamma\gamma$
collisions.  R\&D on this option should be vigorously pursued
and the option to add such collisions after the initial \eebar operation
should be retained.

\vskip 3mm
\leftline{\it 2.3 Low energy collisions }
\vskip 1mm

We have noted that in some physics scenarios we would benefit
from substantial new samples of $Z$ bosons.  These could of course
be obtained at loss of peak luminosity by reduction of the beam energies.
Recent implementations
of the NLC have investigated an IR layout in which the full energy 
beams are brought with little or no bend to a collision point (thus 
facilitating later upgrades of the energy of the collider);  a lower
energy beam can be envisioned that is picked off early in the linacs
and accelerated `for free' using unused portions of the power cycle.
Because of the permanent magnets foreseen for focussing the linac
beams, only about a factor of two variability of beam energy
would be possible on the short term.   Thus, there is the potential
for two ranges of collision energy:  from 0.5 -- 1.0 times
full energy, and from the Z-pole to half the full energy (up to perhaps
a maximum of 500 GeV).

A Giga-$Z$ sample, obtainable with 30 \ifb of data, should make
substantial improvements in the precision of many EW parameters
\cite{moenig}.  Factors of 15 improvement are foreseen for
$\sin\theta^2_W$ and $A_b$.   Factors of 2 -- 5 improvement
should be possible for $\Gamma(Z\rightarrow \ell\ell)$ and $R_b$.
100 \ifb of data at the $WW$ threshold could improve the $W$ 
mass error to below 10 MeV.   Taken together, these should
improve the knowledge of radiative corrections (the $S$ and $T$ parameters)
by about a factor of 8, to the point where extremely tight constraints
can be placed on potential models for new physics.  It is likely that
ultimately one would want to obtain the Giga-$Z$ sample;  in the case
that we do not have manifest new physics at the LHC and LC ({\it e.g.}
supersymmetry) it will be imperative to obtain these data.
It remains an open question for discussion 
whether an optimum program has distinct
experiments focussed on the upper and lower ranges of energies.

\vskip 3mm
\leftline{\it 2.4 $e^- e^-$ collisions }
\vskip 1mm

It should be straightforward to provide $e^-e^-$ collisions at the
LC, though the luminosity will be somewhat reduced relative to \eebar
owing to the absence of the self focussing effects.    Some supersymmetry
studies, searches for new phenomena such as lepton compositeness,
flavor mixing effects ($\widetilde e \rightarrow \mu \widetilde \chi^0$,
strong $WW$ scattering, or searches for large extra dimensions are
enabled or enhanced by $e^-e^-$ collisions.   The nature of new physics
found in the initial \eebar running will indicate how useful the $e^-e^-$
operation might be, but clearly one would not want to preclude such
running in the design phase of the machine.

\vskip 3mm
\leftline{\it 2.5 Free electron laser physics }
\vskip 1mm

Beams of very short wavelength, extremely high peak brightness, photons
are desirable for a broad range of physics, chemistry, biology and material
science applications.  Beams based upon free electron lasers would 
dramatically increase the scope of synchrotron light applications
beyond the present third generation light sources.   The full range
of applications are only now beginning to be understood \cite{arthur}. 
They include the study of `warm plasmas' such as found in planetary
interiors or ion beams; very high-field atomic physics exploring
exotic atomic states and non-linear effects;  nanoscale dynamics
in condensed matter involving short time correlations and collective effects;
femtosecond probes of chemical reactions; and perhaps most exciting,
the possibility of structural studies of biomolecules such as protein
complexes with angstrom level resolution.

In the US, the LCLS 
project is proposed to build a self-amplified spontaneous
emission (SASE) free electron laser based on the last third of the SLAC
linac.  The TESLA project has incorporated an ambitious FEL component
using electron beams of up to 30 GeV, transported to special experimental
halls to the side of the high energy physics IRs.   The LCLS could come
into first operation in 2006, and will begin to map out the range of
experiments that are feasible in the new very high brightness regime.
A linear collider project starting sometime later could build on
the experience of this pilot project, and provide facilities for
a broad international community to explore a wide range of studies.
We should nurture the connection between the LC project and this
broader scientific community to promote new opportunities for 
structural studies of matter.  The joint interests of the two communities
should interfere positively to give a more compelling argument for
the LC project.

\vskip 3mm
\leftline{\it 2.6 Detector issues }
\vskip 1mm

Work has continued in the past two years to define the scope
of linear collider detectors, to conduct R\&D on specific
detector technologies, and to simulate performances for
proposed subdetectors.   Recent developments
for detector simulations\cite{graf}, vertex detectors\cite{brau},
tracking detectors\cite{behnke}, calorimetry\cite{brient},
muon detectors\cite{piccolo}, data acquisition\cite{ledu},
and machine interface\cite{markiewicz} are summarized 
elsewhere in the proceedings.

We assume that there will be two detectors operating in the LC;
for most, the need for independent confirmation
of new discoveries and important measurements dictates
this.  The broad outlines of the LC detectors
have been understood for some time, based on the performance
of existing collider detectors and the general goals of the LC
physics program.  Each linear collider proposal will be accompanied
by general detector designs, sufficient to estimate their scope and
cost.  However, it is wise to defer specific
design choices and optimizations until such time as real detector
collaborations are established, and the full range of interconnected
optimizations can be made by those responsible for building
the detectors.

In the meantime, there are some interesting design choices that
emerge, and these deserve more attention for simulation and 
understanding the physics needs.    The choices of vertex detector
technology at present include CCDs and variants of active silicon
pixel devices.   It is clear that the best affordable vertex
detector will be needed for the LC physics program, with the main
constraints coming from $b,c$ tagging ({\it e.g.} Higgs branching
ratio measurements, $t\overline t H$ coupling, Susy $AH^0$ production, etc.).
Since the need for the best performance vertex detector is clear,
emphasis is needed on the R\&D program to develop commercially viable
detectors, packaging and readout structures.

The other major detector issue concerns calorimetry. 
For physics studies such as Higgs pair production in association
with a $Z$ boson (to measure the Higgs self couplings)
there is a premium on the best possible jet energy
resolution.  For a light Higgs, the decay is dominantly
$H\rightarrow b \overline b$.   The $Z \rightarrow q \overline q$
decay is preferrable due the higher branching ratio.  Disentangling
the jet combinatorics and suppressing backgrounds depends directly
upon the jet energy (and dijet mass) resolution.   A study \cite{gay}
showed that as the jet energy resolution improved from 60\%/$\sqrt E$
to 30\%/$\sqrt E$, the background level decreased by a factor of 6
and the precision on the di-Higgs cross section improved by a factor
of 1.6.  Studies of an `energy-flow' calorimeter have been conducted
in which a very finely segmented calorimeter with small Moli\`ere
radius (small transverse shower spread) is used to identify clusters
caused by charged tracks, and replace their calorimeter energy deposit
with the corresponding charged track measurement.  Other advantages
of such a finely segmented calorimeter may include particle identification
algorithms for $\tau$s and photons.    The cost of the finely divided
calorimeter is however large.  The TESLA detector envisions active layers
of silicon pads (about 1 cm$^{-2}$) 
with tungsten absorber plates.  

It is essential to
further refine the studies of the energy-flow calorimeter with full
simulations and tests to verify the performance characteristics,
and to explore the tradeoffs further.
An added consideration is the extent to which the radius of such a calorimeter
may be reduced to help control costs;  this will in turn place added
burdens on the interior tracking region detectors.

\vskip 6mm
\leftline{\large {\it 3. Scenarios for new physics at the linear collider}}
\vskip 1mm
 
At present we do not have a clear understanding of the Higgs sector and
electroweak symmetry breaking, so the detailed plan for experimentation
at the LC and the need for upgrades to the accelerator
are not wholly understood.  We can however describe several possible
scenarios for the way that physics could play out.  It is important
to develop representative examples, both to frame the 
LC proposals and to assure ourselves that in all imaginable scenarios
the LC has a crucial role to play.   The charge to this workshop\cite{komamiya}
and the recent discussion of the 500 GeV program\cite{uswhite} began
the examination of such scenarios.  We briefly review some of them
here, realizing that fleshing these out is an important role for
the planning exercises now underway in each region.

\begin{enumerate} 
\item
{\sl There is a Higgs below 130 GeV and evidence for supersymmetry
has been found at the Tevatron/LHC.}

The highest initial priority
will be to measure the Higgs properties thoroughly at 500 GeV or below.  
The character of the Susy states accessible at 500 GeV should be determined.

The energy upgrade to about 1 TeV will surely be needed.  
Measuring the Higgs Yukawa couplings to the top quark requires 700 -- 800 GeV.
The remaining 
sparticles and heavy Higgs must be observed and studied.   The full exploration
of the Susy breaking sector will likely require the study of the more
massive gauginos and sfermions.

\item
{\sl There is a Higgs boson below 180 GeV, but no evidence for supersymmetry.}

Again, the Higgs boson parameters must be fully measured at 500 GeV; high
precision is desired since these parameters may hold crucial clues
on the nature of physics beyond the SM.   

Substantial operation at the $Z$ pole, and $WW$, \ttbar ~thresholds will
be desired
since the precision constraints on non-SM models can substantially
limit possible new theories.   The $WWV$ and $ttZ$ anomalous couplings
should be measured with high precision as these are also indicators
of possible new physics.

Later operation at the highest available energy is likely needed,
to study the anomalous $ttH$ couplings, to seek deviations of $WW$
scattering from SM EW production, to look for evidence
of large extra dimensions, etc.

\item
{\sl There is a Higgs boson between 180 -- 300 GeV, 
and no evidence for supersymmetry.}

We note that in this variant of scenario 2, 
we have departed from the SM since the
Higgs mass now exceeds the precision measurements limit.

At 500 GeV, we will want to measure as much as we can about the Higgs,
but the fermionic branching ratios are likely inaccessible.  
(This would be disappointing since one of the hallmarks of the Higgs
is its coupling to fermion mass;  however in this scenario none of
the proposed new colliders could make these measurements.)
The Higgs
quantum numbers, total width, possible CP-violating effects and the
Higgs self-couplings remain critical and achievable 
goals for the 500 GeV program.

The need for precision measurements and for ultimate increase in the energy
are similar to those in scenario 2.

\item
{\sl There is no  Higgs boson and no evidence for supersymmetry at the LHC.}

At 500 GeV, there remain loopholes to close.  A possible Higgs with
invisible decays can be sought.  The anomalous $WWV$ and $ttZ$ couplings
must be measured.

A return to the $Z$ pole and $WW$ threshold will be needed since
in this scenario, we are casting in the dark for hints on the new physics
and the precision \eebar measurements will be crucial.

\item
{\sl We have a Higgs, evidence for supersymmetry, and other
new physics signatures all superimposed.}

The world is so complex that the precision given
by the linear collider will be essential for disentangling the 
new physics.   The LC and LHC with their complementary strengths
will both play essential roles, and their programs will remain
viable for years.

\end{enumerate} 

Some of the scenarios are better developed, and some have larger
sets of measurements that we can envision.   But, at least
in my view, there are none for which the LC is not needed,
and none for which there are not needed measurements at $/sim$500 GeV or
below.  Even in the case that we see little new (e.g. only a Higgs boson
or nothing new at all), we still have to understand why the SM seems
to work so well despite its many theoretical shortcomings.

\vskip 6mm
\leftline{\large {\it 4. Issues for the international high energy physics 
community}}
\vskip 1mm
  
The decisions on a linear collider are likely to be made over the next
few years.   The TESLA proposal will be submitted to the German
government in early 2001, and a recommendation may be 
expected roughly a year after.  The JLC proposal is advancing,
with work on milestones, sites and costs now underway.  The R\&D program
for the NLC is laid out for the next three years, with a proposal
thought feasible in 2003 -- 2004.

All three regions are currently engaged in studies
of long-term ($\sim$20 years) physics issues, and the range of facilities that
might be proposed to address them.  These studies should be complete
in about a year.   In addition to a TeV scale linear collider,
these discussions address other possible projects: a muon storage ring/neutrino
source, a muon collider, a multi-TeV two-beam linear \eebar collider, a very
large hadron collider and new large underground laboratories.
According to a recent HEPAP review\cite{hepap}, none of the alternates
to the LC is expected to be ready for a technical proposal before
the decade beginning in 2010.
In addition, CERN has begun to evaluate its program after
the LHC is operational.  In addition to the projects listed above, CERN
might consider substantial upgrades to the LHC energy or luminosity.

The following comments reflect some personal views on how we may
approach a decision on the LC.

\begin{enumerate}
\item
{\sl Should the LC be the next world high energy machine?}

I believe that it is inevitable that the LC decision will be the next
to be taken by the world community.   Real proposals are 
being made and these will be considered in the next few years.  
There are no proposals for alternate colliders that can be
made on the time scale for LC consideration.
Of course, it is not
necessary that all regions propose a LC in their region,
or even propose substantial financial engagement.  But each region
will have to make a decision on how -- or whether -- 
to address the LC issue soon.  A region may opt out of the 
LC process, but will this will not alter the worldwide timetable for
decision.
This means that the physics planning and priorities activities
now underway in each region have very special urgency, to bring
some coherent community view of the future facilities we need..

We should expect that no more than one LC will be built worldwide.
Gaining approval for the project in any region will be enhanced by support from
all regions.   There could be an argument to forego engagement with
a LC proposals in some region in favor of some other project,
but there is little historical precedent to suggest that such a strategy
would enhance the later, alternative project.

\item
{\sl Is the linear collider too expensive?}
The fate of the SSC and the 1999 cost estimate of the NLC has led
some, particularly in the US, 
to worry that a LC proposal will have difficulties in gaining
government approval.
This is of course a generic problem, since it is likely that the 
cost of any new high energy collider will have a multi-billion dollar
cost.   We will not hide from this problem by substitution of another
project in place of the LC.

Some worry that the cost is the major reason for an initial first
phase at 500 GeV, and that the physics needs may be insufficiently addressed
in that first phase.    I believe that in the past few years, we have
come to a qualitatively new understanding that there is excellent
physics justification for the $~$500 GeV machine.  As indicated
in the scenarios discussion, this now seems more independent of the
specific discoveries at the LHC.  The recent precision measurements
have made it much less likely that the first indications of new physics
would occur at a scale much above 500 GeV.    The studies of supersymmetry seem
assured to be rich at 500 GeV.   Alternate models of new physics seem
to have some observable consequences at 500 GeV.  Making these arguments
for the 500 GeV first phase to the broad HEP community 
is an important responsibility of the proponents of a LC.

The cost of the LC is a factor, and any ways to control the cost
are worth pursuit.   However, the inevitablity of future need for 
higher energy should dictate that we design the upgradability into
the initial stage.

\item
{\sl Where will the LC be?}
There have been numerous comments that there will be only one LC
and that all regions should support its construction at any site.
In practice, most would strongly prefer that it be built in one's
own region.     However, we must realize that a final decision
will be taken with not only scientific considerations in play.
A major factor will be which region is willing to pay the largest
share of the cost.

If we are to have a viable LC program, and to retain scientific 
health in all regions, it will be imperative that the LC is a worldwide
collaboration, both for the accelerator and for the experiments.  Each
region needs some frontier high energy collider activity to keep its
program healthy.  This can be enhanced in part through true 
inter-regional collaboration on each major new facility.  

For the overall health of the community, it is preferrable to
avoid putting most of the contemporaneous frontier facilities in one region.

\item
{\sl How can international collaboration on accelerator projects be achieved?}

Internationalism will mean that compromises are necessary;  although 
a facility will be 
sited in one region, other regions should undertake major responsibilities
in its design, construction, and operation.   This is in part necessary
to keep a healthy community of accelerator scientists in each region.
So, I imagine that for the LC, major subsystems should be taken 
by each region as primary
responsibilities from start to finish.   The global accelerator concept
put forth by ICFA is a valuable start in defining the process by which
this inter-regional collaboration could occur.   Much work remains to
make this model possible.  Accelerator projects are intrinsically more 
tightly controlled and managed than detector projects, but the experience
with international detectors is valuable.  So also are the international
contributions to the LHC accelerator, 
but what is needed to give each region a truly
crucial stake in the LC is beyond what was attempted at the LHC.

\item
{\sl Technical evaluations of LC proposals}

The panel discussion of some Lab directors at this workshop highlighted
a proposal for a technical review of all LC proposals before approval
or site decisions are made.    Such a review is imagined to start around
the end of 2001.   It would focus on the technical solutions for all phases of
the project, risks and needed R\&D.  It should probably address costs in
some common currency.   Upgrade paths would be useful to examine.
The technical review process should not address
specific site issues, or  attempt to make final decisions on the preferred
technology beyond assessment of risk or cost factors.  The technical review
process should be seen as an important set of considerations that will
inform the physics community and guide governmental bodies as they undertake
to make a decision on approving and siting a collider.
Details of how this process would work remain to be worked out.
What body would charge the review panel, choose its members and receive
its report?   What timescale should be adopted for its activities
that will not place any particular region at disadvantage technically
or politically?  But the benefit to the world HEP community and those
who need to understand the technological and cost issues from this
review are large, and we should encourage the formation of this process.

\end{enumerate}


\begin{references}
\bibitem{moenig}
K. M\"onig, ``What is the Case for a Return to the $Z$-Pole?'', 
these proceedings.
\end{references}
\end{document}
